\newcommand*{\wn}{cm$^{-1}$}

\documentclass[aps,prl,twocolumn,groupedaddress]{revtex4}
\usepackage{graphicx}
\usepackage{verbatim}
\usepackage{subfigure}
\usepackage[english]{babel}
\usepackage{verbatim}
\usepackage{multirow}

\begin{document}

\title{Dissociation energy of the hydrogen molecule at 10$^{-9}$ accuracy}

\author{C. Cheng, J. Hussels, M. Niu, H. L. Bethlem, K. S. E. Eikema, E. J. Salumbides, W. Ubachs}
\email{w.m.g.ubachs@vu.nl (W. Ubachs)}
\affiliation{Department of Physics and Astronomy, LaserLaB, Vrije Universiteit Amsterdam, de Boelelaan 1081, 1081 HV Amsterdam, The Netherlands}

\author{M. Beyer, N. J. H\"{o}lsch, J. A. Agner, F. Merkt}
\affiliation{Laboratorium f\"{u}r Physikalische Chemie, ETH Z\"{u}rich, 8093 Z\"{u}rich, Switzerland}

\author{L.-G. Tao, S.-M. Hu}
\affiliation{Hefei National Laboratory for Physical Sciences at Microscale, iChem center, University of Science and Technology China, Hefei, 230026 China}

\author{Ch. Jungen}
\affiliation{Laboratoire Aim\'{e} Cotton du CNRS, B\^{a}timent 505, Universit\'{e} de Paris-Sud, F-91405 Orsay, France}
\affiliation{Department of Physics and Astronomy, University College London, London WC1E 6BT, United Kingdom}

\date{\today}

\begin{abstract}
The ionization energy of ortho-H$_2$ has been determined to be $E^\mathrm{o}_\mathrm{I}(\mathrm{H}_2)/(hc)=124\,357.238\,062(25)$ \wn\ from measurements of the GK(1,1)--X(0,1) interval by Doppler-free two-photon spectroscopy using a narrow band 179-nm laser source and the ionization energy of the GK(1,1) state by continuous-wave near-infrared laser spectroscopy. $E^\mathrm{o}_\mathrm{I}$(H$_2$) was used to derive the dissociation energy of H$_2$, $D^{N=1}_{0}$(H$_2$), at $35\,999.582\,894(25)$ \wn\ with a precision that is more than one order of magnitude better than all previous results. The new result challenges calculations of this quantity and represents a benchmark value for future relativistic and QED calculations of molecular energies.
\end{abstract}


\maketitle

The dissociation energy of the hydrogen molecule $D_0$(H$_2$) is a fundamental quantity for testing molecular quantum theory. The pioneering calculations of Heitler and London on H$_2$ demonstrated that molecular binding is a consequence of quantum mechanics \cite{Heitler1927}. In almost a century of mutually stimulating activities by experimentalists and theorists, the accuracy of this benchmark value has been improved by seven orders of magnitude \cite{Sprecher2011}. Great progress in the theoretical calculations has been achieved by including relativistic and quantum-electrodynamic (QED) effects \cite{Wolniewicz1995,Piszczatowski2009}. Over the past decade, improved calculations of the Born-Oppenheimer energies \cite{Pachucki2010}, adiabatic corrections \cite{Pachucki2014}, leading-order nonadiabatic corrections \cite{Pachucki2015}, exact nonadiabatic energies \cite{Pachucki2016,Simmen2013}, and a further refinement of QED calculations \cite{Puchalski2016} have been reported. The latest efforts, however, led to a deterioration of the agreement between experimental and theoretical results \cite{Puchalski2017}.

\begin{figure}
    \includegraphics[width=3.2in]{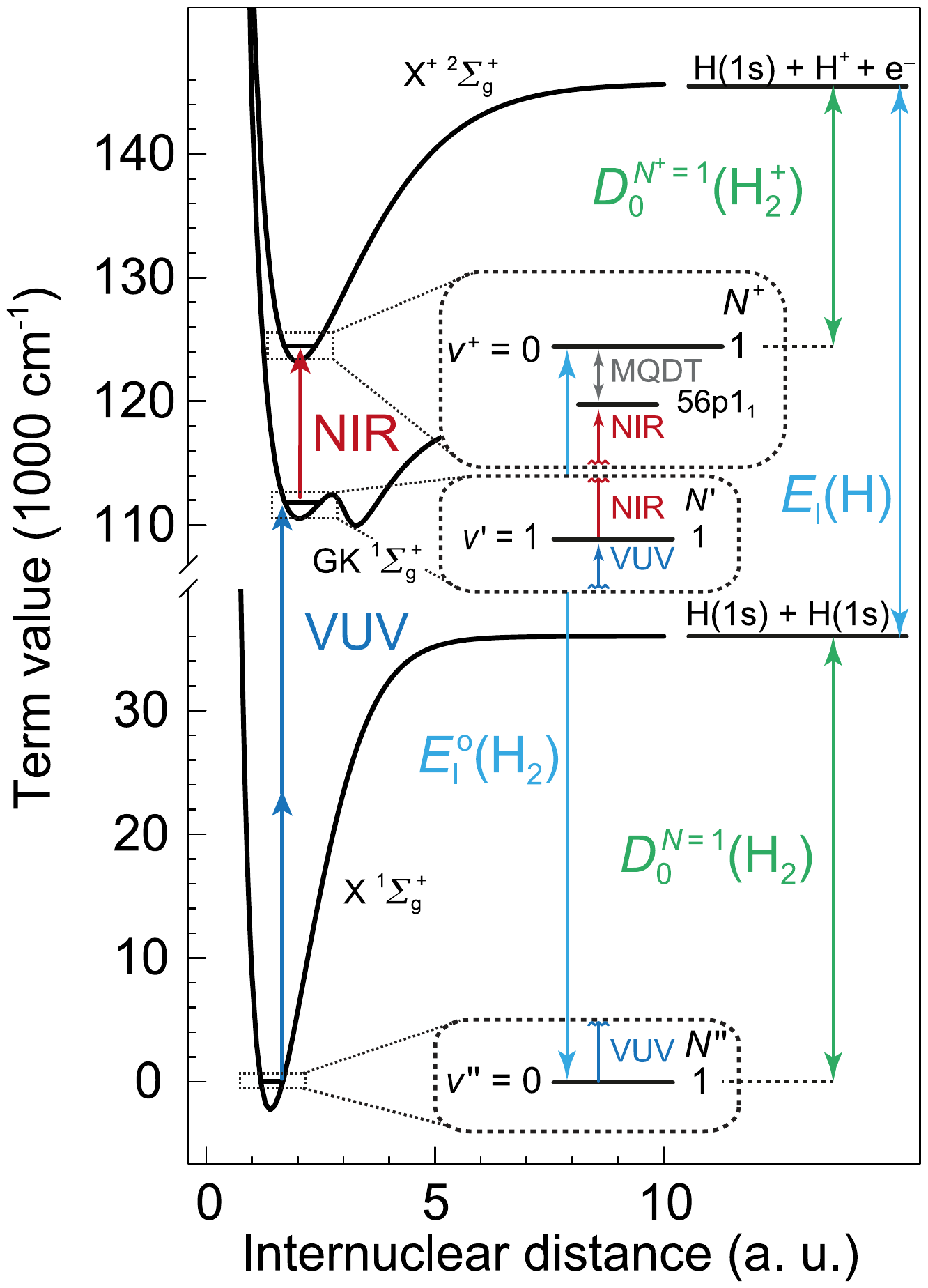}		
    \caption{(Color online) Potential energy diagram of electronic states in molecular hydrogen relevant to this study.
            \label{Fig_Potential}
    }
\end{figure}

Direct measurements of dissociation energies in H$_2$ are complicated by perturbing resonances near the continuum limits and vanishing direct photo-dissociation cross sections at the thresholds \cite{Herzberg1970,Zhang2004}. Such difficulties can be overcome by a measurement of the adiabatic ionization energy $E_\mathrm{I}$(H$_2$) combined with a thermodynamic cycle involving the ionization energy of atomic hydrogen $E_\mathrm{I}$(H) and the dissociation energy of the molecular ion $D_\mathrm{0}$(H$_2^+$) \cite{Herzberg1972} via
\begin{displaymath}
D_\mathrm{0}(\mathrm{H}_2) = E_\mathrm{I}(\mathrm{H}_2) + D_\mathrm{0}(\mathrm{H}_2^+) - E_\mathrm{I}(\mathrm{H}),
\end{displaymath}
as illustrated in Fig. \ref{Fig_Potential}. The dissociation energies $D^{N=1}_{0}$ and $D^{N=0}_{0}$ of ortho- and para-H$_2$ differ by the rotational term value of the X($v=0,N=1$) level, i.e., $118.486\,84(10)$ \wn\ \cite{Jennings1984}.

The most accurate previous determination of $D_0$(H$_2$), at a relative accuracy of $10^{-8}$ \cite{Liu2009}, involved two-photon Doppler-free laser excitation to the EF $^1\Sigma^+_g$($v=0,N=1$) intermediate state \cite{Hannemann2006}, one-photon ultraviolet excitation from the EF $^1\Sigma^+_g$(0,1) to the $56$p1$_1$ Rydberg state \cite{Liu2009}, and millimeter wave (MMW) spectroscopy of high-lying Rydberg states \cite{Osterwalder2004} allowing for an extrapolation to the ionization energy by Multi-Channel Quantum Defect theory (MQDT) \cite{Haase2015}. The initial EF-X step in this scheme has recently been improved by two orders of magnitude to an accuracy of 73 kHz \cite{Altmann2018}, but an improvement of $D_0$(H$_2$) awaits an improved measurement of the EF-$n$p interval.

In the present work, we adopt an alternative excitation scheme to determine $D_0$(H$_2$), through the GK $^1\Sigma^+_g$(1,1) intermediate state, which offers the possibility of using continuous wave (cw) infrared laser excitation to high-$n$ Rydberg states \cite{Beyer2018}. Experimental results from two laboratories are combined: the measurement of the Doppler-free two-photon transition $\text{GK}(1,1) \leftarrow \text{X}(0,1)$ in Amsterdam, and the determination of the interval between GK(1,1) and the $56$p1$_1$ Rydberg state by near-infrared (NIR) cw-laser spectroscopy in Z\"{u}rich.

\begin{figure}
    \includegraphics[width=3.2in]{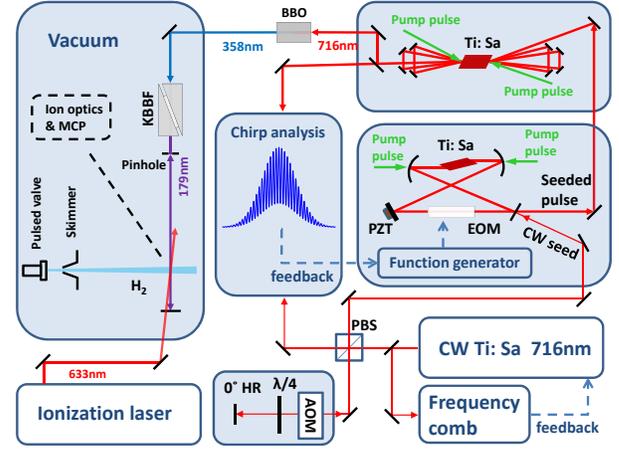}
    \caption{(Color online) Schematic layout of the GK-X experimental setup, where the pulsed pump laser for the oscillator and amplifier is not shown.
            \label{Fig_Setup}
    }
\end{figure}

In the GK-X experiment, schematically depicted in Fig. \ref{Fig_Setup}, a narrow bandwidth ($\sim$9 MHz) injection-seeded oscillator-amplifier titanium sapphire (Ti:Sa) laser system delivers 50-ns-long pulses at the fundamental wavelength of 716 nm. The amplified pulsed output is frequency up-converted in two doubling stages, with BBO and KBe$_{2}$BO$_{3}$F$_{2}$ (KBBF) crystals, leading to the generation of 179-nm radiation to drive the $\text{GK}(1,1) \leftarrow \text{X}(0,1)$ transition in a two-photon scheme. The vacuum-ultraviolet (VUV) output power of 20 $\mu$J per pulse is limited by the optical damage threshold of the KBBF crystal \cite{Zhang2008}. A glass pinhole with a diameter of 0.5 mm is employed to align the reflected beam in a counter-propagating Doppler-free configuration. A separate 633-nm pulsed dye laser is used to ionize the molecules from the GK(1,1) state in a single-photon ionization process. To reduce AC Stark effects, this laser is delayed by 30 ns with respect to the 179-nm pulse. Further increase of the delay is detrimental because the lifetime of the GK(1,1) state is 24(3) ns \cite{Astashkevich2015}. The H$_2^+$ ions are collected and detected by the velocity-map-imaging method \cite{Eppink1997}.

The cw Ti:Sa laser, which has a short-term (one second) frequency stability of a few tens of kHz, is locked to an optical frequency comb, resulting in a long-term relative accuracy better than 10$^{-12}$. An acousto-optic modulator (AOM) was implemented in a double-pass scheme to scan the cw laser frequency, the output of which is used as a seed for the oscillator cavity. The frequency offset or chirp, between the pulsed output and cw seed was measured for each pulse \cite{Hannemann2006}. An intra-cavity electro-optic phase modulator (EOM), driven by an arbitrary function generator, is used for active frequency chirp compensation, as shown in Fig. \ref{Fig_Setup}. In addition, the residual chirp value is recorded and corrected at each frequency scan step. An upper limit of the systematic uncertainty associated with chirp is extracted from the statistical analysis, by repeating the measurements after changing the anti-chirp parameters.

\begin{figure}
    \includegraphics[width=3.5in]{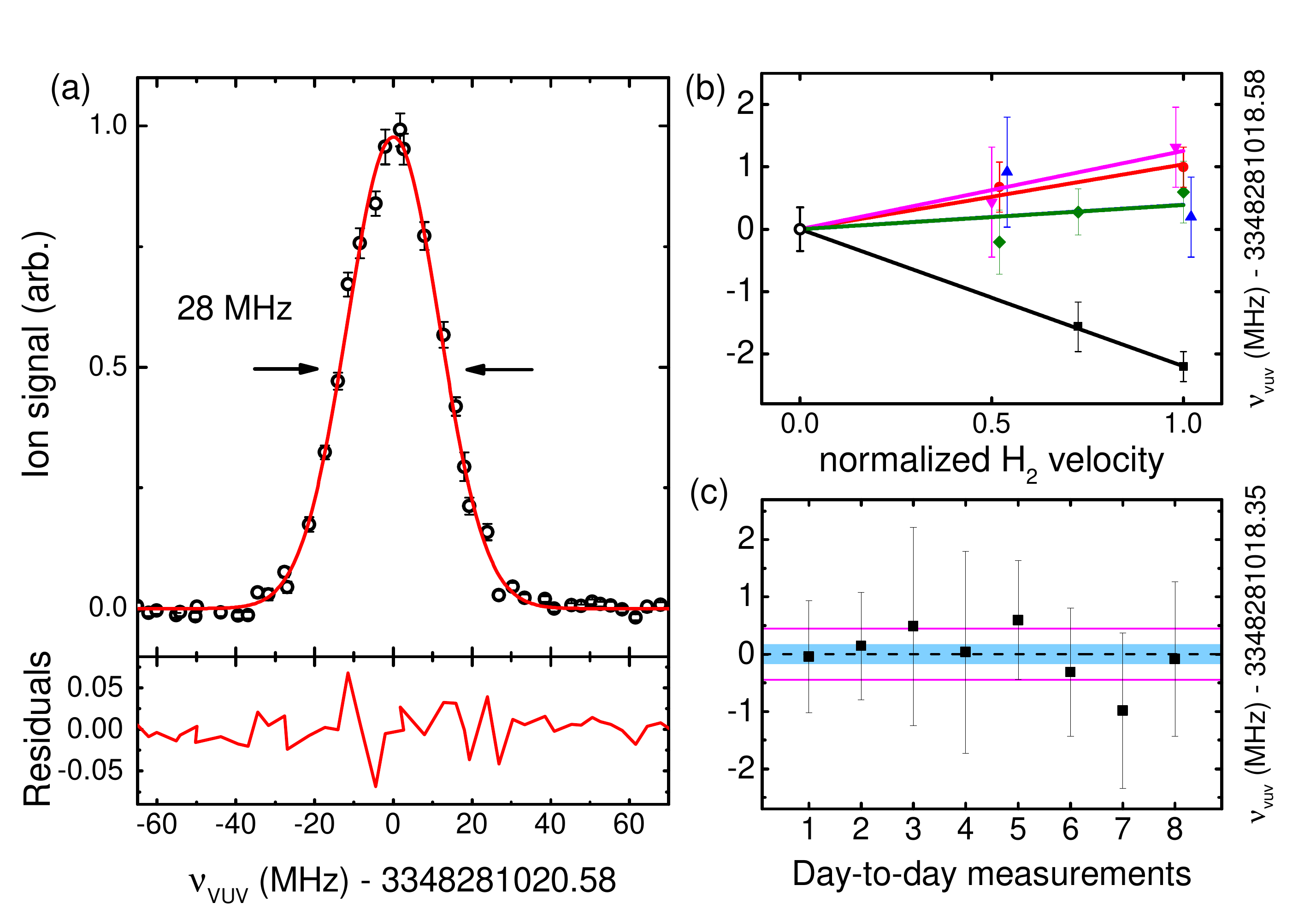}
    \caption{(Color online) (a)~Recording of a $\text{GK}(1,1) \leftarrow \text{X}(0,1)$ transition in H$_2$ (black circles). The red line is a fitted Gaussian curve with the residuals shown below. (b)~Assessment of the residual first-order Doppler effect. Each color indicates an individual alignment configuration, where the largest deliberate misalignment is about 0.3 mrad. The blue and magenta points are shifted by 0.02 and -0.02 in velocity axis for clarity. The colored lines show a linear global fit for all alignments, resulting in the Doppler-free value indicated with the open circle. (c)~Transition frequency measurements in different days. Each point indicates an average value for one day with its standard deviation. The dash line shows the mean value. The magenta line and the cyan area give the standard deviation of the data and of the mean, respectively, where the former is taken as a conservative statistical uncertainty.
            \label{Fig_Results}
    }
\end{figure}

A typical scan is shown in Fig. \ref{Fig_Results}(a) with 50-shot averaging for each frequency scan step. The observed two-photon transition linewidth is dominated by the laser bandwidth, with a small contribution of the natural linewidth ($\Gamma=6.6$ MHz). An imperfect counter-propagating alignment may result in a residual first-order Doppler shift. This was quantified by performing velocity-dependent measurements using various mixtures of H$_2$ and Ne, and extrapolating to a zero-velocity transition frequency, as shown in Fig. \ref{Fig_Results}(b). Several measurements were performed using different alignment configurations of the counter-propagating VUV laser beams. After accounting for the second-order Doppler shift, which is 150(30) kHz in pure H$_2$ with a velocity of 2900(300) m/s, a global fitting procedure is applied, where the zero-velocity intercept is shared for all alignment settings. The extrapolation yields the Doppler-free transition frequency with a systematic uncertainty of 350 kHz, which is the largest contribution to the error budget. The normalized velocity of the H$_2$ beam in Fig. \ref{Fig_Results} is defined as $v_{\mathrm{norm}}=v_{\mathrm{mix}}/v_{\mathrm{pure}}=\sqrt{m_{\mathrm{H_2}}/(n_{\mathrm{H_2}}\cdot m_{\mathrm{H_2}}+n_{\mathrm{Ne}}\cdot m_{\mathrm{Ne}})}$, where $n_{\mathrm{H_2}}$ and $n_{\mathrm{Ne}}$ are the mixture fractions of H$_2$ and Ne, and $m_{\mathrm{H_2}}$ and $m_{\mathrm{Ne}}$ are their masses \cite{bookScoles1988}.

The AC Stark effect for both the 179-nm and the ionization lasers was studied by performing intensity-dependent measurements. Typically the 179-nm laser power was fixed to 2 $\mu$J per pulse during the residual Doppler-shift determination, while up to 10 $\mu$J was generated to assess the AC Stark effect. A similar procedure is applied for the ionization laser, including the assessment of systematic shifts caused by the temporal overlap between the two laser pulses. The Doppler-extrapolated value was corrected for the AC Stark shifts. Other possible systematic and statistical uncertainties were derived from day-to-day frequency differences (in total 215 measurements) over several days (see Fig. \ref{Fig_Results}(c)). The uncertainty budget is given in Table~\ref{tab:uncertainties}, and the combined statistical and systematic uncertainty of the GK-X transition is 650 kHz, corresponding to a relative accuracy of 2$\times$10$^{-10}$.

\begin{table*}
\centering
\caption{Transition frequencies of H$_2$ and their uncertainties.}
	\begin{tabular}{lcccccc}
	\hline
	\hline
	\textbf{Transition} &$\hspace{0.5cm}$&\multicolumn{2}{c}{$\mathbf{\text{GK}(1,1) \leftarrow \text{X}(0,1)}$}&$\hspace{0.5cm}$ &\multicolumn{2}{c}{$\mathbf{56\text{p}1_1 \leftarrow \text{GK}(1,1)}$} \\
	Measured frequency 		& &\multicolumn{2}{c}{$3\,348\,281\,018.58(49)$~MHz}& &\multicolumn{2}{c}{$378\,809\,479.24(30)$~MHz} \\
	\cline{1-1}\cline{3-4}\cline{6-7}
	Effect                                    &          & Correction & Uncertainty &            & Correction & Uncertainty \\
    \cline{1-1}\cline{3-4}\cline{6-7}
    DC Stark shift					                 &      &          & $<$10~kHz                        &  &  & 7~kHz\\
	AC Stark shift					                 &      & -40~kHz  & 90~kHz,$^a$                      &  &  & 4~kHz\\
                                                     &      & -190~kHz & 200~kHz,$^b$                     &  &  &     \\
    Chirp						                     &      &          & ($<$490~kHz)$_{\text{stat}}$,$^c$&  & --  & \\
	Zeeman shift 					                 &      &          & $<$10~kHz                        &  &  & 10~kHz\\
	Collision shift				                 &      &          & $<$1~kHz                         &  &  & 1~kHz\\
	Residual first-order Doppler shift		         &      &          &  350~kHz                         &  &  & ($<$110~kHz)$_{\text{stat}}$,$^c$ \\
	Second-order Doppler shift		                 &      &          & $<$30~kHz,$^d$                   &  & +4.1~kHz& 0.5~kHz\\
	Line-shape Model                       		     &      & --       &                                  &  &  & 200~kHz\\
	Hyperfine structure (c.g. shift)          	     &      &          & $<$100~kHz                       &  &  & 100~kHz\\
    Photon-recoil-shift correction		             &      & --       &                                  & & -160~kHz& \\
	\cline{1-1}\cline{3-4}\cline{6-7}	
	Total systematic uncertainty		             &      &          & 426~kHz                          &  &  & 224~kHz \\
	Final frequency 				                 &  &\multicolumn{2}{c}{$3\,348\,281\,018.35(49)_{\text{stat}}(43)_{\text{sys}}\,\text{MHz}$} & &\multicolumn{2}{c}{$378\,809\,479.08(30)_{\text{stat}}(22)_{\text{sys}}\,{\text{MHz}}$} \\
	\hline
    \hline
	\end{tabular}
\flushleft
$^a$ For the ionization laser.
$^b$ For the VUV laser. \\
$^c$ This systematic uncertainty is already included in the statistical uncertainty of the frequency measurements.\\
$^d$ The second-order Doppler shift values are subtracted for different velocities in Fig. ~\ref{Fig_Results}(b) and the error is included in the residual first-order Doppler shift uncertainty.
\label{tab:uncertainties}
\end{table*}

\begin{figure}[h]
   \includegraphics[width=0.95\linewidth]{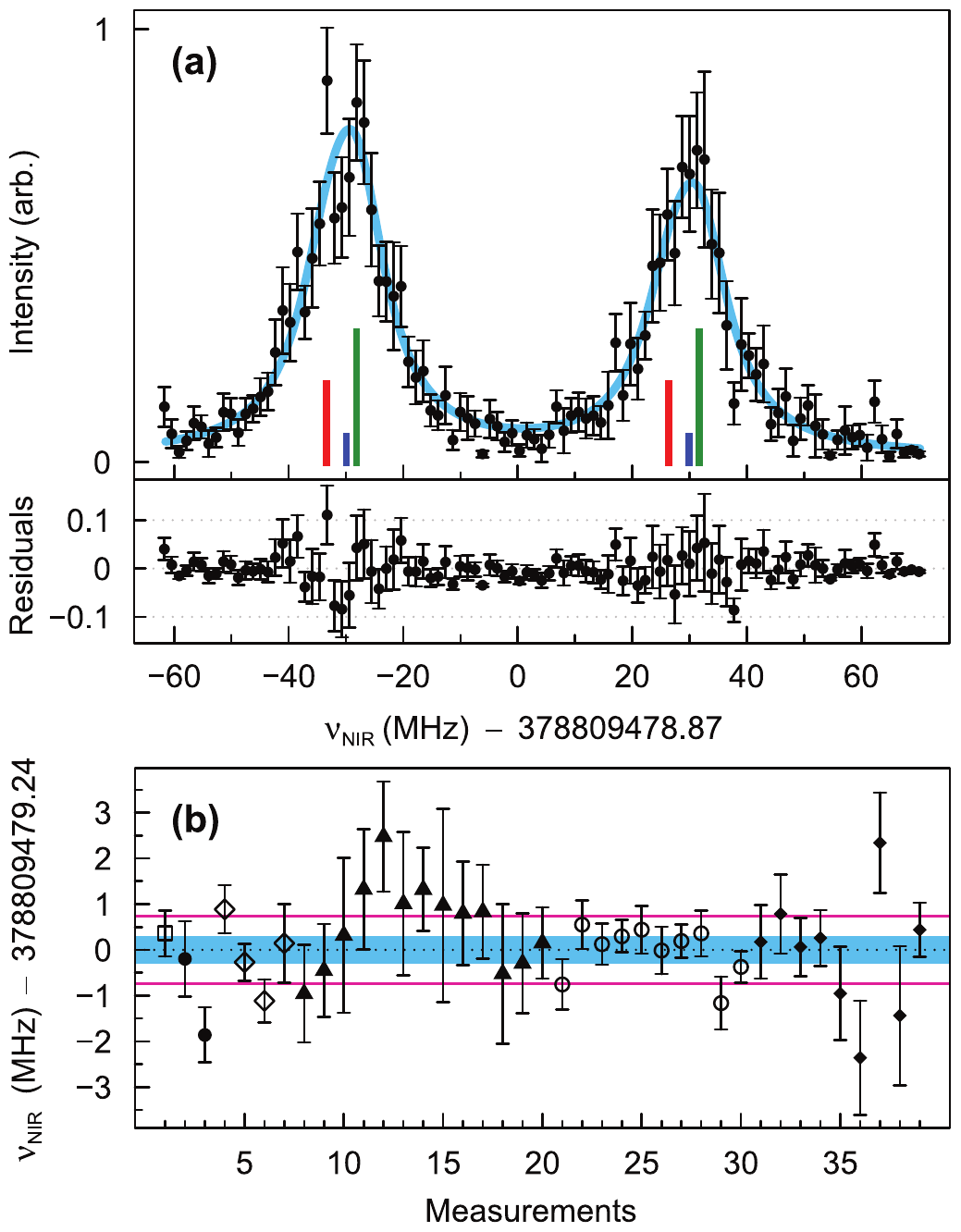}
   \caption{(a) Typical spectrum of the $56\text{p}1_1 \leftarrow \text{GK}(1,1)$ transition of H$_2$ and its analysis based on a Lorentzian line-shape model and the hyperfine structure of the 56p1$_1$ Rydberg state given as red ($F=1$), blue ($F=0$) and green ($F=2$) sticks. The weighted residuals are depicted below the spectrum. (b) Doppler-free frequencies with standard deviations of individual measurements. The magenta line and the cyan area give the standard deviation of the full data set and of the mean, respectively. The symbols label measurements carried out on different days.
\label{Fig_RydbergSpec}
}
\end{figure}

The interval between the GK(1,1) state and the 56p$1_1(v^+=0,S=0,F=0-2)$ Rydberg state of ortho-H$_2$ was measured using the same apparatus, laser setup and calibration procedure as described in detail in a recent article presenting a measurement of the $50\text{f}0_3 \leftarrow \text{GK}(0,2)$ interval \cite{Beyer2018}. The measurement was carried out using a pulsed and skimmed supersonic beam of pure H$_2$ and the procedure involved:

(i) the compensation of the stray electric fields in three dimensions to better than 1~mV/cm, which limits possible Stark shifts to below 7~kHz for the 56p$1_1$ level \cite{Sprecher2013},

(ii) shielding external magnetic fields so that the maximal Zeeman shifts are below 10~kHz,

(iii) the cancellation of the first-order Doppler shift to better than 110~kHz by performing the excitation with the NIR-laser beam of 792-nm wavelength and its back reflection overlapped to better than 0.05 mrad and averaging the central frequencies of both Doppler components (see Fig. \ref{Fig_RydbergSpec}(a)); repeating the measurements after full alignment of the laser and molecular beams transforms the systematic uncertainty associated with the residual Doppler shift into a statistical uncertainty,

(iv) cooling the valve used to generate the supersonic beam to 80~K, thus reducing the mean beam velocity to 1290(20)~m/s and leading to a second-order Doppler shift of -4.1(5)~kHz,

(v) calibrating the excitation frequency with a frequency comb referenced to a 10-MHz Rb oscillator (Stanford Research Systems, FS275).

A typical individual spectrum is displayed in Fig.~\ref{Fig_RydbergSpec}(a) as dots with error bars. The figure also depicts a fit of a model line shape consisting of two Doppler components. Each of these is the sum of three hyperfine components (stick spectrum) having the same Lorentzian width, intensities given by their statistical weights of $2F+1$, and relative positions corresponding to those measured by millimeter-wave spectroscopy \cite{Osterwalder2004, Sprecher2011}. In the fit, the data points were weighted by taking into account the Poissonian statistics of the ion counts and the background noise as explained in Ref. \cite{Beyer2018}. The central positions determined from 39 measurements are plotted in Fig.~\ref{Fig_RydbergSpec}(b) with their statistical uncertainties. Their weighted mean, indicated by the dashed line, is $378\,809\,479.24(30)$~MHz where the statistical uncertainty of 300~kHz corresponds to $\bar{\sigma}=\sigma/\sqrt{N}$, $N$ being the number of independent measurements, which we took to be the number of measurement sets recorded on different days rather than the number of individual measurements (i.e., 6, as indicated by the different symbols in Fig.~\ref{Fig_RydbergSpec}(b), rather than 39).

The systematic uncertainties considered in the analysis are summarized in Table~\ref{tab:uncertainties} and sum up to 224~kHz, dominated by the uncertainty resulting from the possible deviations from a statistical intensity distribution of the unresolved hyperfine structure of the $56\text{p}1_1 \leftarrow \text{GK}(1,1)$ line (Line-shape Model entry in Table \ref{tab:uncertainties}). After subtraction of the photon recoil shift of 160~kHz, our final result for the $56\text{p}1_1 \leftarrow \text{GK}(1,1)$ interval is shown in Table \ref{tab:uncertainties}. This value is consistent with, but four times more precise than, the value of $378\,809\,478.7(12)$~MHz reported in \cite{Sprecher2013}.

\begin{table*}
\caption{Energy level intervals and determination of the ionization $E_\mathrm{I}$ and dissociation energies $D_0$ of ortho-H$_2$ (in cm$^{-1}$).
 \label{Table_Ei}
}

\begin{tabular}{c@{\hspace{0.3cm}}c@{\hspace{0.3cm}}r@{.}l@{\hspace{0.3cm}}c@{\hspace{0.3cm}}c}
\hline%
\hline%
 & Energy level interval &\multicolumn{2}{c}{Value $\hspace{15pt}$} & Ref. & Comment \\
\hline%
(1)&GK$(v=1,N=1)$ -- X$(v=0,N=1)$           & 111\,686&632\,836(22)      & This work &\\
(2)&56p$1_1(v^+=0,S=0,{\text{center}})$ -- GK$(v=1,N=1)$      & 12\,635&724\,114(12)       & This work &\\
(3)&X$^+(v^+=0,N^+=1,{\text{center}})$ -- 56p$1_1(v^+=0,S=0,{\text{center}})$  & 34&881\,112(5)         &  \cite{Sprecher2014} &\\
(4)&[H(1s) + H$^+$] -- X$^+(v^+=0,N^+=1,{\text{center}})$     & 21\,321&116\,575\,5(6)     &  \cite{Korobov2017,CODATA2014} &$D^{N^+=1}_0$(H$_2^+$)\\
(5)&[H(1s) + H$^+$] -- [H(1s) + H(1s)]          &   109\,678&771\,743\,07(10) & \cite{CODATA2014} &$E_\mathrm{I}$(H)\\
(6)&(1)+(2)+(3)                                 &124\,357&238\,062(25)& This work &$E^{\text{o}}_\mathrm{I}$(H$_2$)\\
(7)&(1)+(2)+(3)+(4)-(5)                     & 35\,999&582\,894(25)& This work &$D^{N=1}_0$(H$_2$)\\
\hline%
\hline%
\end{tabular}
\end{table*}

The experimental values of the $56\text{p}1_1 \leftarrow \text{GK}(1,1)$ interval in Table \ref{tab:uncertainties} correspond to the center of gravity (c.g.) of the hyperfine components. The hyperfine splitting of the GK(1,1) state, which has d character, is estimated to be 330 kHz from the known hyperfine structure of high-$n$d Rydberg states \cite{Osterwalder2004} and leads to a systematic uncertainty contribution of 100 kHz for the transition center frequency. In addition, the hyperfine splitting of the X(0,1) state, which was observed by Ramsey to be 600 kHz \cite{Ramsey1952}, also contributes to the systematic uncertainty for the $\text{GK}(1,1) \leftarrow \text{X}(0,1)$ measurement. In the center-of-gravity transition frequency determination, a contribution of less than 100 kHz is estimated.

The binding energy of the $56{\text{p}}1_1$ Rydberg state with respect to the first rovibronic state X$^+$~($v^+=0, N^+=1$) of ortho-H$_2^+$ was determined via a MQDT-assisted fitting procedure applied to 76 measured $n$p hyperfine components with $54<n<64$, as described in Ref. \cite{Sprecher2014} and the value is given in Table~\ref{Table_Ei}. Combining all contributions, the ionization energy of ortho-H$_2$, $E^\mathrm{o}_\mathrm{I}$(H$_2$), is determined to be $124\,357.238\,062(25)$ \wn\ (see Table~\ref{Table_Ei}), corresponding to a relative accuracy of $2\times10^{-10}$. The dissociation energy, $D_0^{N=1}$(H$_2$), is derived from $E^\mathrm{o}_\mathrm{I}$(H$_2$) to be $35\,999.582\,894(25)$ \wn\ or $1\,079\,240\,344.3(8)$ MHz with a relative accuracy of $7\times10^{-10}$, by using the values of $D_0^{N^{+}=1}$(H$_2^+$), calculated to an accuracy of $6\times10^{-7}$ \wn\ \cite{Korobov2017,CODATA2014}, and $E_\mathrm{I}$(H), which is included in CODATA 2014 \cite{CODATA2014}.

\begin{figure}
    \includegraphics[width=3.5in]{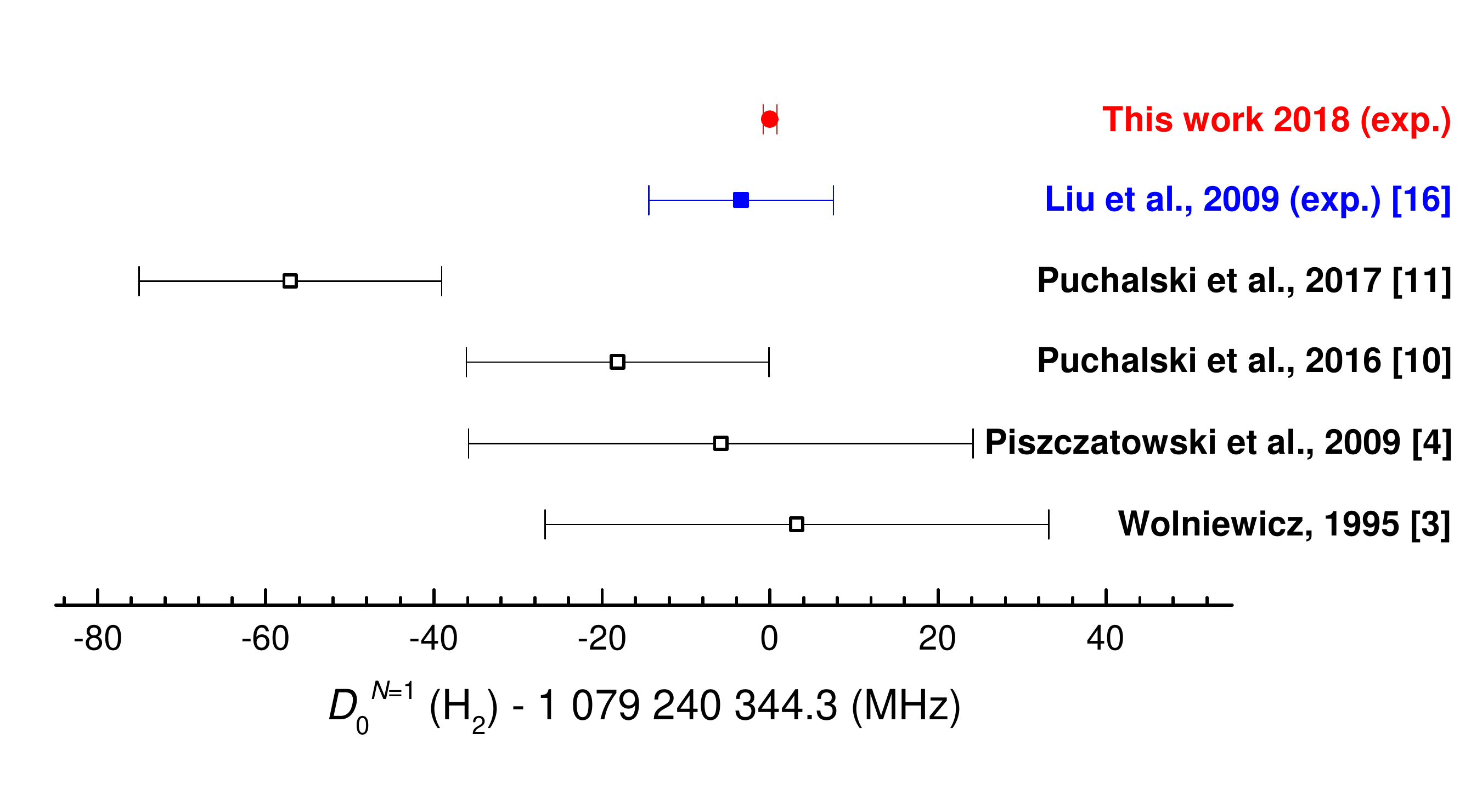}
    \caption{(Color online) Comparison between experimental and theoretical values of $D_0^{N=1}$ for ortho-H$_2$.
            \label{Fig_D0}
    }
\end{figure}

A comparison between our new value of $D_0^{N=1}$(H$_2$) and the most recent experimental and theoretical results is presented in Fig. \ref{Fig_D0}. Our result confirms the validity of the previous experimental result \cite{Liu2009} using a different excitation sequence, but improves its accuracy by one order of magnitude. It deviates from the newest theoretical result reported in Ref. \cite{Puchalski2017} by more than three times the uncertainty.
Possible reasons for the discrepancy between the experimental and theoretical values of $D_0$(H$_2$) include the underestimation of nonadiabatic effects in the determination of the relativistic and QED corrections to $D_0$(H$_2$) \cite{Puchalski2017}, or a more fundamental problem in the molecular quantum theory. Resolving this puzzle and further improvement of this value to 10-kHz accuracy, for both experiment and theory, will open a new route for determining the proton charge radius \cite{Pohl2010,Puchalski2017} with 1\% accuracy, or an improved value of the proton-to-electron mass ratio \cite{Ubachs2016,Tao2018}.

\begin{acknowledgments}
FM and WU acknowledge the European Research Council for an ERC-Advanced grant under the European Union's Horizon 2020 research and innovation programme (grant agreement No 670168 and No 743121). HB, KE and WU acknowledge FOM/NWO for a program grant on ``The Mysterious size of the proton''. FM acknowledges the Swiss National Science Foundation (project Nr. 200020-172620). SH acknowledges the support from NSFC (21688102) and CAS (XDB21020100).
\end{acknowledgments}

\bibliography{Hydrogen}


\end{document}